\documentclass[aps,physrev,twocolumn,groupedaddress]{revtex4-2}
\usepackage{graphicx}
\usepackage{amsmath}
\usepackage{dcolumn}
\usepackage{bm}

\begin{document}


\title{The application of annealing in quantum cooling protocols.}


\author{Xu Chongyuan}
\email[]{Moke2001@whu.edu.cn}
\affiliation{}


\date{\today}

\begin{abstract}
Inspired by simulated annealing algorithm, we propose a quantum cooling protocol which includes an annealing process. This protocol can be universally and efficiently applied to various quantum simulators, driving the system from an arbitrary initial state to the ground state with high fidelity. We have described the cooling process based on perturbation theory, validated the advantages of bath under time-modulated Zeeman field compared to bath under static one, and provided a justification for the necessity of an annealing process when the system to be cooled is unknown. We applied tensor network methods to numerically simulate our cooling protocol, using the transverse field Ising model (TFIM) as an example to verify the effectiveness of the protocol in cooling one-dimensional systems, two-dimensional systems, and systems with quantum noise. We compared the overall performance of cooling protocols with and without the annealing process on a test set generated with random parameters $g_P$. The results indicate that the cooling protocol with annealing process can achieve both accuracy and efficiency. Our results also show that the cooling protocol's resistance to noise depends on the type of quantum noise.
\end{abstract}


\maketitle

\section{Introduction}

The preparation of the ground state of quantum many-body systems is a key task in quantum computing and quantum simulation\cite{poulin2009preparing}. Quantum simulation needs the system's ground state to provide important information about its characteristics, including interactions, phase transitions, and collective behaviors\cite{georgescu2014quantum}. By studying the ground state, we can gain the effective models of complex quantum systems and observe novel quantum phenomena such as quantum spin liquids\cite{zhou2017quantum} and the quantum Hall effect\cite{cage2012quantum}. Quantum computing needs the correct preparation of the ground state to promise the success of many quantum algorithms. In quantum annealing algorithms, problems are encoded into a Hamiltonian, and by preparing the ground state of this Hamiltonian, we can obtain solutions.\cite{rajak2023quantum}.

Previously, the methods for preparing the ground state of many-body systems were mainly through variational quantum circuits\cite{cerezo2021variational}\cite{khairy2020learning}\cite{du2022quantum} and adiabatic evolution\cite{farhi2000quantum}\cite{roland2002quantum}\cite{farhi2001quantum}. However, both of them face varying degrees of difficulties. Variational quantum algorithms are a quantum-classical hybrid algorithm based on variational optimization, suitable for medium-scale noisy quantum circuits. Optimizing quantum circuits based on deep learning, which is to achieve approximate solutions to relevant problems without full error correction. However, the optimization process of variational quantum circuits can become very complex due to high-dimensional parameter space, and "barren plateau" often renders this methods ineffective\cite{mcclean2018barren}\cite{larocca2024review}. Adiabatic evolution is a state preparation method based on the quantum adiabatic theorem, but this approach cannot overcome the effects of quantum noise and relies on the choice of paths. As soon as the ground state and the first excited state cross, the adiabatic evolution method fails\cite{roland2005noise}\cite{comparat2009general}.

Mimic cooling is a scheme newly proposed for preparing the ground state of many-body systems, aiming at a quantum many-body system $P$ with a Hamiltonian $\hat H_P$\cite{boykin2002algorithmic}\cite{kaplan2017ground}\cite{feng2022quantum}. We introduce a bath $A$ with a Hamiltonian $\hat H_A$, where the ground state of $\hat H_A$ is known and diagonalized in the experimental representation. We can easily prepare its ground state $|E_0^A\rangle$ and directly observe the excited states of $A$. There exists a weak interaction $\lambda\hat H_{AP}$ between $A$ and $P$. $A$ and $P$ combine to form a total system, whose Hamiltonian is:

\begin{equation}
\hat H=\hat H_P+\hat H_A+\hat H_{AP}    
\end{equation}

Cooling protocol includes three basic steps: evolution, measurement, and reset. We first prepare system $P$ in state $|\psi_0^S\rangle$, evolve $A$ and $P$ under $\hat H$ for a time $T$, then we obtain a non-product state entangling $A$ and $P$. A measurement is performed on $A$ to reset the qubits that are excited, bringing $A$ back to its ground state. At this point, the state vector of whole system becomes $|\psi_1^PE_0^A\rangle$. We repeat these three steps until no excitations are measured from $A$ for a long duration. As shown in Figure 1, we illustrate the configuration of the system and bath used in this paper, as well as the basic process of cooling protocol.

\begin{figure}[htbp]
\centering
\includegraphics[width=\linewidth]{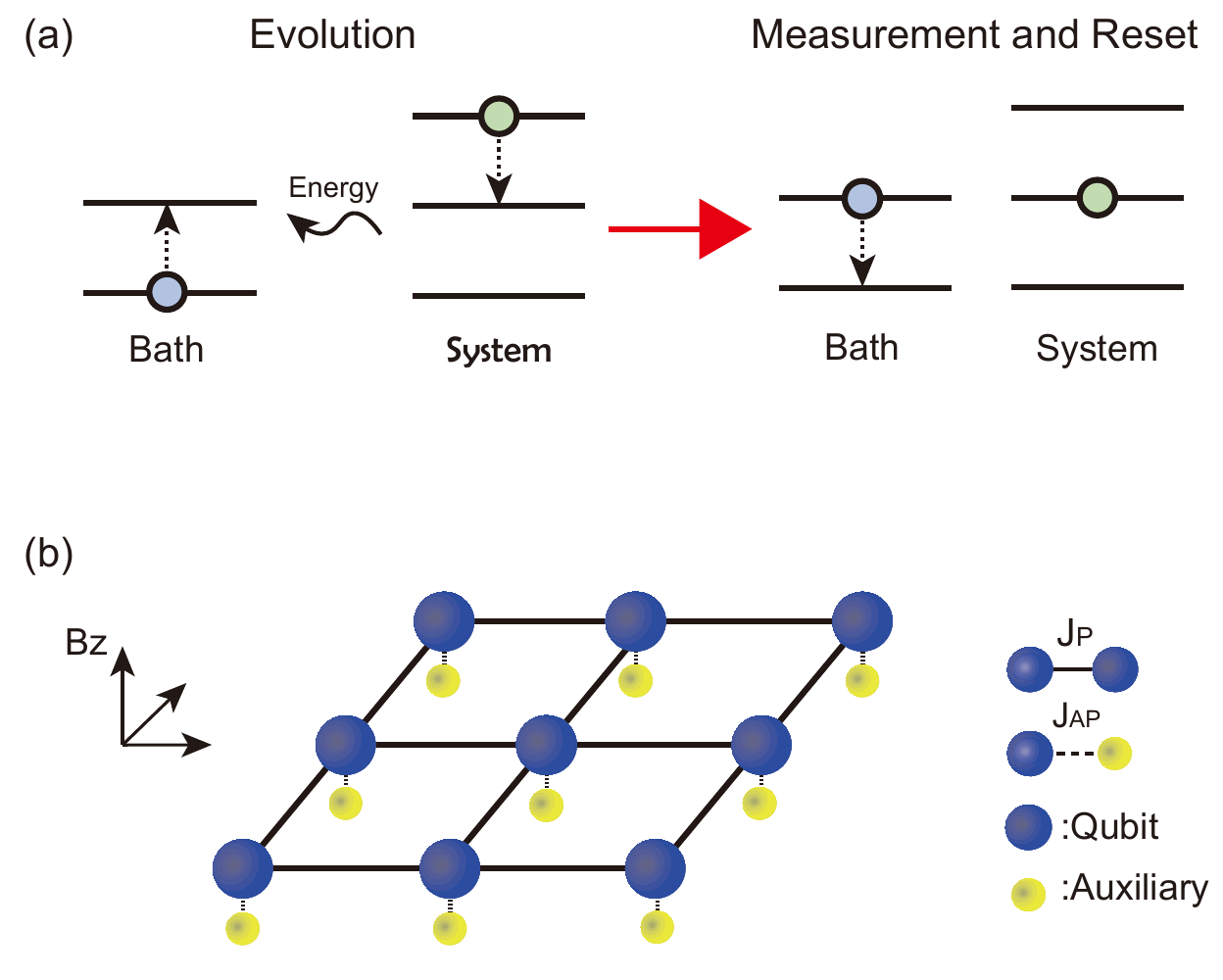}
\caption{(a) First, the system and the bath undergo a unitary evolution, while heat and entropy being transferred from the system to the bath. Subsequently, through measurement and resetting of the qubits in the bath, heat and entropy are extracted. (b) The system studied in this paper is a one- or two-dimensional atomic system, where each qubit in the system interacts with an auxiliary qubit, which is the bath. There are nearest-neighbor interactions between the qubits in the system, and different external magnetic fields are applied to the system and the bath.}
\end{figure} 

Mimic cooling has made significant advancements in both theory and experiment in prior research. Mi et al. implemented an experiment using a cooling protocol to cool a quantum system, demonstrating that the cooling protocol can efficiently cool one-dimensional systems but inefficiently be used in two-dimensional systems\cite{mi2024stable}. Anne et al. proposed using a bath under time-modulated Zeeman field to cool quantum systems, which provides greater adaptability to the cooling protocol and mitigates the efficiency drop due to differences in the excitation spectra of systems\cite{matthies2022programmable}. Based on these research, this paper presents a perturbation theory that describes the cooling process and proves that time-modulated bath have greater advantages compared to static ones. As we face diverse and complex systems, it is necessary to introduce an annealing process into the cooling protocol in cases where there is no priori knowledge of the Hamiltonian. We provide a justification for this necessity. To validate the correctness of the theory and the effectiveness of protocol, we utilized tensor networks to simulate the cooling protocol's implementation, which is verified on the one-dimensional TFIM, the two-dimensional TFIM, and the TFIM containing noise. Additionally, we generated a test set of cooling protocols with random parameters $g_P$, comparing the overall performance of cooling protocols with and without the annealing process, further confirming that the cooling protocol incorporating the annealing process can achieve both accuracy and efficiency.

The remaining sections of this paper will unfold as follows: In the second section, we propose our cooling protocol; in the third section, we provide our analysis based on perturbation theory, including a dynamic description of the cooling process and our justification for introducing annealing; in the fourth section, we present the results of numerical simulations; the final section summarizes and discusses our findings.


\section{Cooling protocols}

Considering a system of an atomic array, $P$ is a one- or two-dimensional plane atomic array with the problem Hamiltonian, namely $\hat H_P$.

There are two approaches to modulating the energy levels of the cooling source: spatial modulation and temporal modulation. The idea of spatial modulation is to set the excitation energy corresponding to different qubits in $A$ to different values. This can be achieved by breaking the system's symmetry through long-range interactions or disorder. The simplest method is to apply different Zeeman fields to the different qubits. The main drawback of spatially modulating the energy spectrum is that it cannot guarantee that the energy level difference between two energy levels of $P$ connected by local interactions exactly matches the excitation energy of the qubits. Therefore, we focus on the bath through temporally modulating the energy spectrum.

During the evolution process, we change the strength of the Zeeman field while also varying the interaction between $P$ and $A$ strength to minimize its impact on the energy spectrum. The total system evolves under the following Hamiltonian:

\begin{equation}
\hat H(t)=\hat H_P+\hat g_A(t)H_A+J_{AP}(t)\hat H_{AP}
\end{equation}

We set the Zeeman field to be a linearly scanned Zeeman field:

\begin{equation}
g(t)=g_{max}+\frac{g_{min}-g_{max}}{T}t
\end{equation}

We initially increase $J_{AP}(t)$ linearly to its maximum over a time $t_0$, maintain it until $t_1$, and then linearly reduce $J_{AP}(t)$ to $0$, completing one scan.

When faced with an unknown system or disorder system, we often cannot determine the energy structure of the Hamiltonian, making it difficult to give a reasonable value for $J_{AP}^{max}$. If $J_{AP}^{max}$ is set too small, the cooling efficiency will decrease sharply; if it is set too large, it will induce a quantum phase transition in the system, leading to convergence results that do not reflect the true ground state of the system. To address this, we introduce an annealing process: we gradually reduce the interaction strength between the bath and the system within a loop, allowing the system to converge to the ground state more accurate and quickly. The annealing rate is denoted as $v$, and we have:

\begin{equation}
J_{AP}^{max}(N)=J_{AP}^{0}\times v^N
\end{equation}

In this equation, $J_{AP}^{0}$ represents the initial interaction strength, and $N$ represents the number of cycles now.


\section{Analysis}

\subsection{Time-independent System}

The Hamiltonian of the total system composed of both can be expressed using equation (1). We can expand the Hamiltonians of the two systems in terms of their energy eigenstates:

\begin{equation} 
\begin{cases} 
\hat H_P=\sum_{i=0}^{N_P-1}E_i^P|E_i^P\rangle\langle E_i^P|\\
\hat H_A=\sum_{j=0}^{N_A-1}E_j^A|E_j^A\rangle\langle E_j^A| 
\end{cases} 
\end{equation}

We assume that at the start of cooling, the initial state of the total system is a product state $|E_{i_0}^PE_0^A\rangle$. In the interaction picture, the state vector and operators of the system evolve simultaneously. We are interested in projectors of the form $|E_{i_1}^PE_{j_1}^A\rangle\langle E_{i_1}^PE_{j_1}^A|$. This operator remains unchanged under the evolution operator $\hat U_0(t)$ derived from $\hat H_P + \hat H_A$, while the state vector is obtained by applying the evolution operator in the interaction picture, $\hat U_I(t)$.

We first consider the time-independent system. The zeroth-order evolution operator of the system is:

\begin{equation}
\hat U_0(t)=\exp(-i\hat H_P t)\exp(-i\hat H_A t)
\end{equation}

By performing perturbative expansion on the evolution operator in the interaction picture, we obtain its expression:

\begin{equation}
\hat U_I(t)=\hat I-i\int_0^t\hat U_0^{-1}(t')\hat H_{AP}\hat U_0(t'){\rm d}t' \end{equation}

What we want to find is the magnitude of the projection component of the initial state on $|E_{i_1}^PE_{j_1}^A\rangle$ after evolving for a time $t$, which is to calculate the expectation value of the state vector on the projector $|E_{i_1}^PE_{j_1}^A\rangle\langle E_{i_1}^PE_{j_1}^A|$:

\begin{equation}
\begin{aligned} 
|\langle E_{i_1}^PE_{j_1}^A|\psi(t)\rangle|^2= \langle E_{i_0}^PE_{0}^A|\hat U_I^{-1}(t)|E_{i_1}^PE_{j_1}^A\rangle\langle E_{i_1}^PE_{j_1}^A|\\
\hat U_I|E_{i_0}^PE_{0}^A\rangle=|\langle E_{i_1}^PE_{j_1}^A|\hat U_I|E_{i_0}^PE_{0}^A\rangle|^2
\end{aligned}
\end{equation}

Here, $\psi(t)$ represents the state vector at time $t$. Based on the above, we compute $\langle E_{i_1}^PE_{j_1}^A|\hat U^{-1}I|E{i_0}^PE_0^A\rangle$:

\begin{equation}
\begin{aligned} 
\langle E_{i_1}^PE_{j_1}^A|\hat U^{-1}I|E{i_0}^PE_0^A\rangle=-i\langle E_{i_1}^PE_{j_1}^A|\int_0^t\exp(i\hat H_P t')\\
\exp(i\hat H_A t')\hat H_{AP}\exp(-i\hat H_P t')\exp(-i\hat H_A t'){\rm d}t'|E_{i_0}^PE_{0}^A\rangle 
\end{aligned}
\end{equation}

We can move the state vectors on both sides into the integral and simplify to obtain:

\begin{equation} 
\langle E_{i_1}^PE_{j_1}^A|\hat U^{-1}I|E{i_0}^PE_0^A\rangle=-iH_{i_0j_0i_1j_1}\frac{1-e^{i\Delta t}}{\Delta E} 
\end{equation}

Here, $\Delta E$ is the energy difference between the two states:

\begin{equation} 
\Delta E=E_{i_1}^P+E_{j_1}^A-E_{i_0}^P-E_{0}^A 
\end{equation}

We can see that for a higher energy state $|E_{i_0}\rangle$ to transition to a lower energy state $|E_{i_1}\rangle$, the resonance condition $\Delta E=0$ must be satisfied; furthermore, the transition matrix element must not be zero. As shown in Figure 2, a necessary condition for a system to cool to its ground state is that there is a "transition path" leading to the ground state for each energy level of the system.

\begin{figure}[htbp]
\centering
\includegraphics[width=\linewidth]{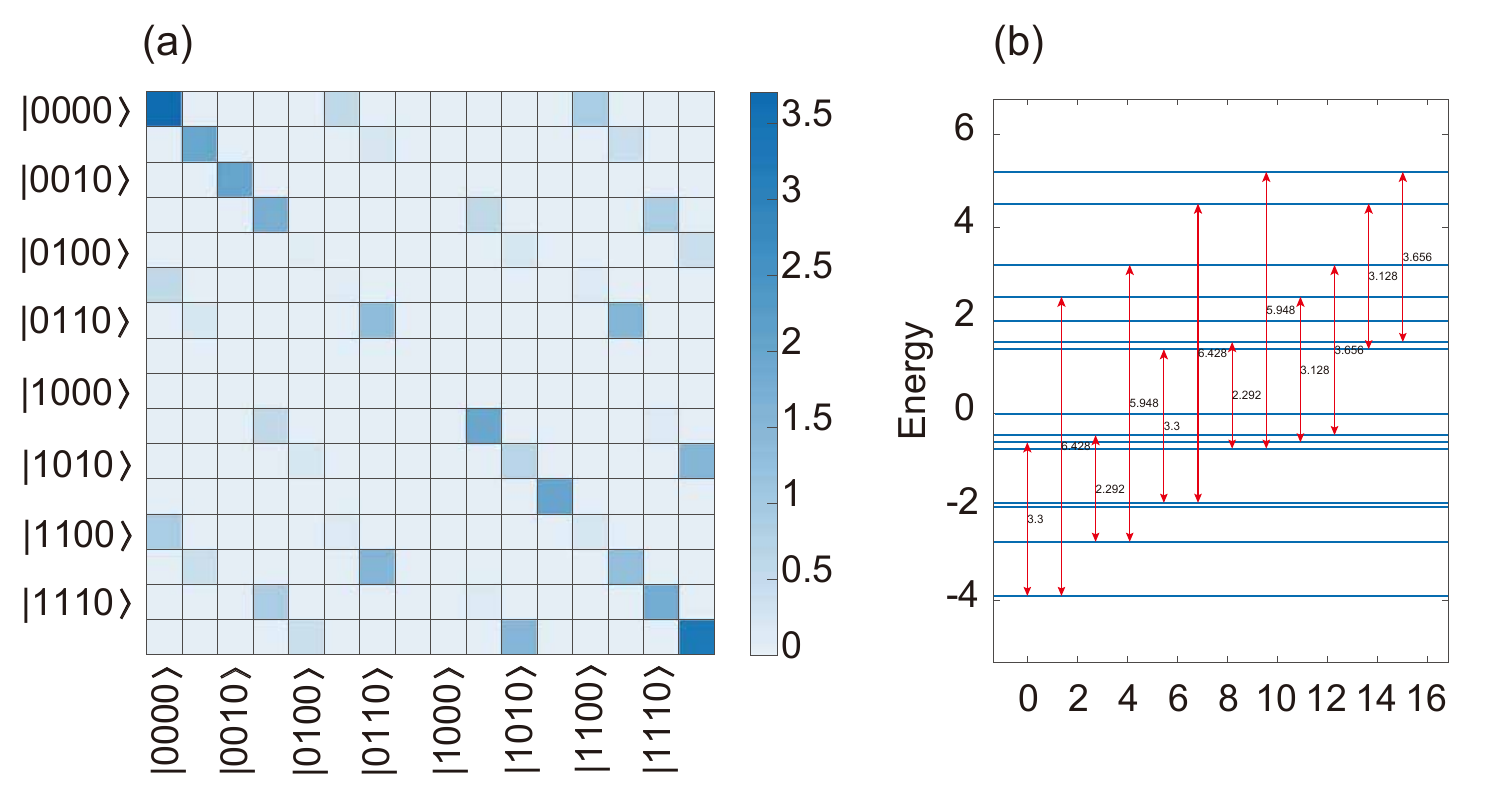} 
\caption{Energy spectrum of the one dimension TFIM, taking $J_P=1, g_P=1.5,N_P=4$, with the interaction with the bath as $\sigma^x_n\sigma^x_m$ interaction. (a) Projection of the transition matrix elements on the eigenstates of the TFIM, where color intensity indicates the transition strength. (b) Energy level structure of the TFIM, with vertical lines marking the energy states that allow a transition.} 
\end{figure}

\subsection{Time-dependent System}

From equation (10), we can see that in a time-independent system, the condition for achieving optimum forward transition is the matching of energy levels, thus a time-independent system cannot ensure that transition moments are always at resonance, which may lead to very slow forward transition rates in some cases. To overcome this difficulty, we utilize a time-dependent Zeeman field to modulate the energy levels of the cooling source, which can, on average, improve the efficiency of transitions.

We denote the Hamiltonian of the time-dependent cooling source system as $\hat H_A(t)=g_A(t)\hat H_A$.

\begin{equation} 
\hat H_A(t)=\sum_{m=0}^{M-1}g_A(t)\sigma_m^z(t) 
\end{equation}

This means that the form of the zeroth-order evolution operator of the system needs to be corrected:

\begin{equation} 
\hat U_0(t)=\exp(-i\hat H_P t)\exp\left(-i\hat H_A\int_0^t g(t'){\rm d}t'\right) \end{equation}

Substituting this into the previous expression for the projection component yields a new expression:

\begin{equation} 
\begin{aligned}
\langle E_{i_1}^PE_{j_1}^A|\hat U^{-1}I|E{i_0}^PE_0^A\rangle=-iH_{i_0j_0i_1j_1}\\
\int_0^t\exp\left(iE_{i_1}^Pt'-iE_{i_0}^Pt'+i\left(E_{j_1}^A-E_{j_0}^A\right)\int_0^{t'}g(t''){\rm d}t''\right){\rm d}t' 
\end{aligned}
\end{equation}

We adopt a linear scan method, setting $g(t)=gt$, along with two parameters:

\begin{equation} 
\begin{cases}
A=\frac{1}{2}g(E_{j_1}^A-E_0^A)\\
B=E_{i_1}^P-iE_{i_0}^P 
\end{cases} 
\end{equation}

The result simplifies to:

\begin{equation}
\begin{aligned}
\langle E_{i_1}^PE_{j_1}^A|\hat U^{-1}I|E{i_0}^PE_0^A\rangle=-iH_{i_0j_0i_1j_1}\\
\int_0^t\exp(i(At'^2+Bt')){\rm d}t'
\end{aligned}
\end{equation}

This equation is a Fresnel integral\cite{umul2005equivalent}, and the integral result is:

\begin{equation}
\begin{aligned}
\langle E_{i_1}^PE_{j_1}^A|\hat U^{-1}_I|E_{i_0}^PE_0^A\rangle=-iH_{i_0j_0i_1j_1}\\
\left(\frac{1}{\sqrt A}\cos(\frac{B^2}{4A})(C(\sqrt At+\frac{B}{2\sqrt A})-C(\frac{B}{2\sqrt A}))+\right.\\
\frac{1}{\sqrt A}\sin(\frac{B^2}{4A})(S(\sqrt At+\frac{B}{2\sqrt A})
-S(\frac{B}{2\sqrt A}))+\\
i\frac{1}{\sqrt A}\cos(\frac{B^2}{4A})(S(\sqrt At+\frac{B}{2\sqrt A})-S(\frac{B}{2\sqrt A}))-\\
\left.i\frac{1}{\sqrt A}\sin(\frac{B^2}{4A})(C(\sqrt At+\frac{B}{2\sqrt A})-C(\frac{B}{2\sqrt A}))
\right)
\end{aligned}
\end{equation}

According to the properties of the Fresnel integral function, as $t\rightarrow\infty$, this projection component approaches a constant. For forward transitions, when $E_{i_1}^P-E_{i_0}^P<0$, both $C\left(\frac{B}{2\sqrt A}\right)$ and $S\left(\frac{B}{2\sqrt A}\right)$ are negative, leading to a relatively large value for the expression. In contrast, the backward transition is suppressed due to $E_{i_1}^P-E_{i_0}^P>0$. In summary, this scheme makes the forward transition of the system easier to occur, outperforming the case with a static Zeeman field.

\subsection{Necessity of Introducing Annealing}

We consider the simplest model, where both the system and the bath are two-level systems with equal energy differences between their two energy levels, and the interaction is of the $\sigma^x_n\sigma^x_m$ form. We assume the interaction strength as $H$, and the energies of the four states $|00\rangle, |01\rangle, |10\rangle, |11\rangle$ are $E_0, E_1, E_1, E_2$, respectively, with the interaction given by:

\begin{equation} 
\hat H_{AP}=H(|00\rangle\langle 11|+|11\rangle\langle 00|+|01\rangle\langle 10|+|10\rangle\langle 01|)
\end{equation}

Diagonalizing the Hamiltonian gives the eigenvectors of the system arranged in a matrix:

\begin{equation}
\begin{array}{l}
\left(\begin{array}{cccc}
0 & 0 & \frac{\frac{E_0 }{2}+\frac{E_2 }{2}-\sigma_1 }{H}-\frac{E_2 }{H} & \frac{\frac{E_0 }{2}+\frac{E_2 }{2}+\sigma_1 }{H}-\frac{E_2 }{H}\\
-1 & 1 & 0 & 0\\
1 & 1 & 0 & 0\\
0 & 0 & 1 & 1
\end{array}\right)\\
\mathrm{}\\
\textrm{where}\\
\mathrm{}\\
\;\;\sigma_1 =\frac{\sqrt{{E_0 }^2 -2\,E_0 \,E_2 +{E_2 }^2 +4\,H^2 }}{2}
\end{array}
\end{equation}

When $H$ is small, we are within the range where first-order perturbation theory is valid. The eigenvectors of the system approximately separate to $|00\rangle$ and $|11\rangle$, and we can consider that there is no transition between the two states. When $H$ is large, the above result does not hold, which implies that cooling cannot maintain the system in the ground state, as there is a strong coupling between the ground state and the excited states $|11\rangle$. However, equation (17) shows that the size of $H$ determines the transition strength,which is proportional to cooling efficiency. Therefore, if $J_{AP}^{max}$ is too small, the cooling efficiency will decrease sharply.

Based on the discussion above, it is natural to introducing annealing into the cooling protocol. When the system is at a higher energy level, there are many coupled levels below, and a larger interaction strength will not significantly affect the cooling trend. Conversely, when the system is overall at a lower energy level, close to the ground state, we must ensure that the transition from the ground state to excited states is kept as small as possible, which corresponds to a smaller interaction strength. Therefore, introducing an annealing process is a natural consideration.


\section{Numerical Simulation}

We consider the Hamiltonian of the problem system is a TFIM, given by:

\begin{equation}
\hat H_P=\sum_{n=0}^{N-1}g_P\sigma_n^x+\sum_{}J_P\sigma_{n_0}^z\sigma_{n_1}^z
\end{equation}

We will test the effectiveness of the protocol on this model.

\subsection{One-Dimensional Transverse Field Ising Model}

We first test the cooling protocol on the one-dimensional TFIM, which has been studied extensively. We set the initial state of the system as a Neel state; of course, one could also set it to other polarized states or thermal states, as there are no restrictions on the cooling protocol regarding the initial state. We use the energy difference value $\frac{E-E_0}{E_0}$ as an indicator of the system's ground state fidelity, as shown in Figure 3. Each gray curve in the figure represents a single cooling process. Due to the measurement process during cooling, we can consider this to represent a quantum trajectory. Because measurement is stochastic, these gray curves exhibit discontinuous change points. By averaging across many samples, we can obtain the ensemble evolution, as indicated by the blue curve in Figure 3.

Figures 3(a) and 3(b) correspond to the cases with and without quantum noise, respectively. The presence of quantum noise significantly alters the system's cooling results, corresponding to the Lindblad operator associated with quantum noise in the master equation, thus changing the steady state accordingly. When the system is free of noise, most samples can smoothly converge to the ground state, with only a very few instances showing sudden energy increases due to measurement-induced uncertainties. After introducing quantum noise, the system may experience energy increases at any time due to the noise.

\begin{figure}[htbp]
\centering
\includegraphics[width=\linewidth]{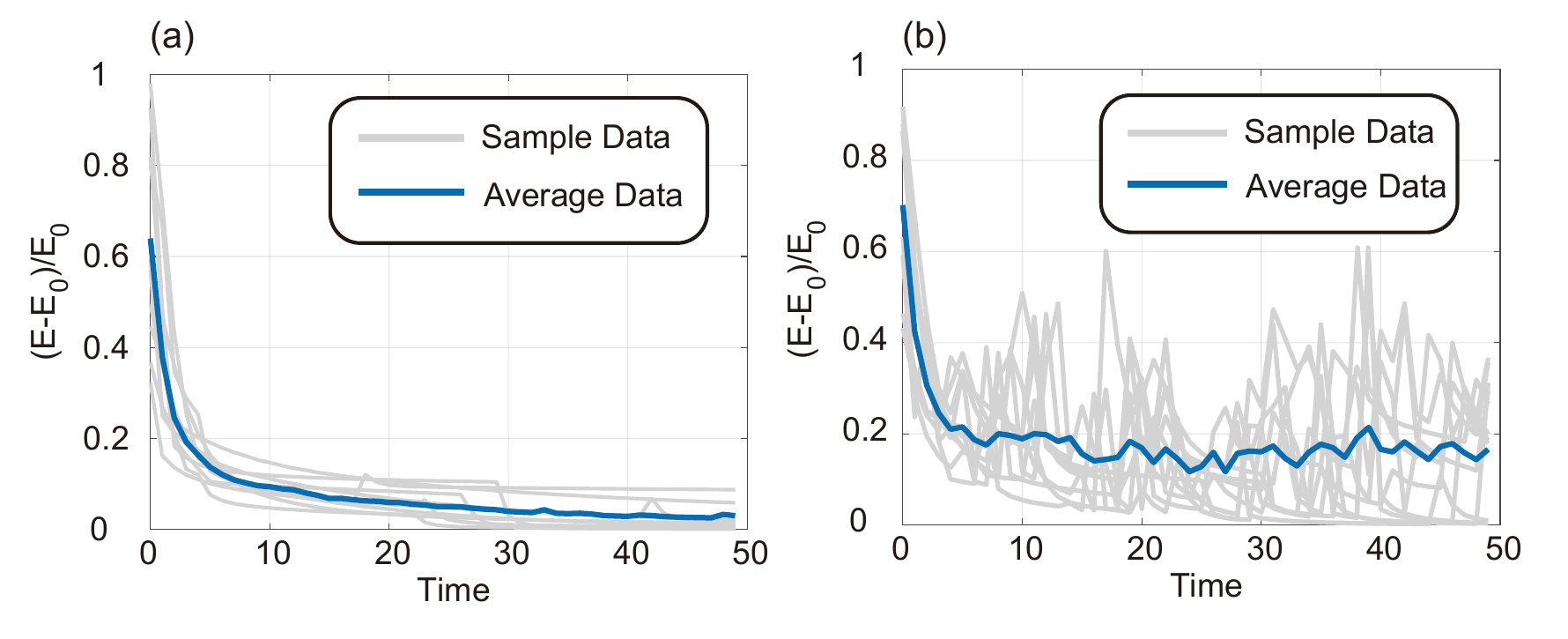}
\caption{Cooling process of the cooling protocol on the one-dimensional TFIM. The horizontal axis represents time, while the vertical axis represents the energy difference from the ground state. The parameters are set to $J_P=1, g_P=1.5, N_P=8$, and the initial state is a Neel state, averaging over 30 samples, with 10 samples displayed as the gray curves. (a) Cooling on the transverse field Ising model without noise. (b) Cooling on the TFIM with noise, where the noise takes the form of a local depolarizing channel with noise strength of $0.01$ on each qubit.}
\end{figure}

\subsection{Two-Dimensional Transverse Field Ising Model}

Now we test the cooling protocol on the two-dimensional TFIM while current studies mainly focus on one-dimensional systems. We verify that the cooling protocol is also effective in two-dimensional systems. We still set the initial state of the system as a Neel state, using a square lattice model of size $3\times3$. As shown in Figure 4, the cooling protocol can effectively cool the system to a high-fidelity ground state, both with and without noise.

\begin{figure}[htbp]
\centering
\includegraphics[width=\linewidth]{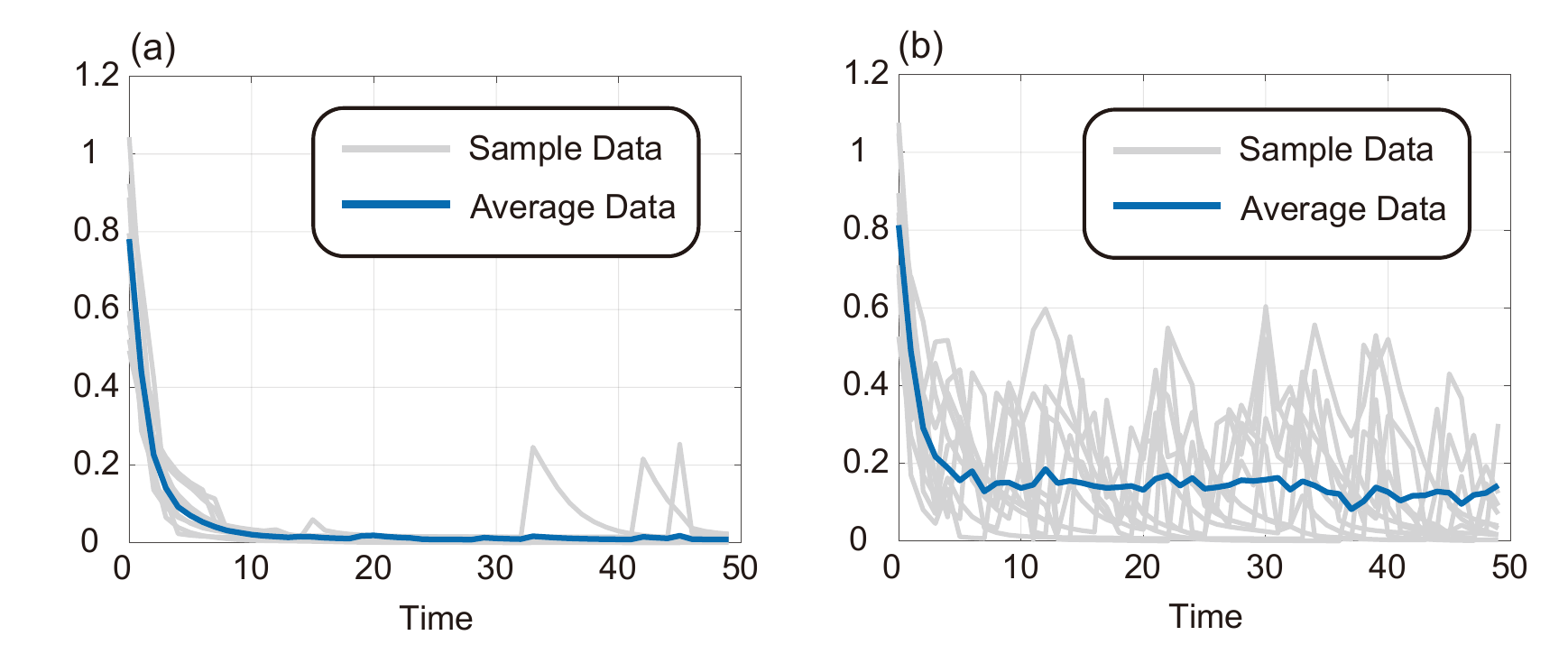}
\caption{Cooling process of the cooling protocol on the two-dimensional TFIM. The horizontal axis represents time, while the vertical axis represents the energy difference from the ground state. The parameters are set to $J_P=1, g_P=1.5, N_P=9$, and the initial state is a Neel state, averaging over 30 samples, with 10 samples displayed by the gray curves. (a) Cooling on the transverse field Ising model without noise. (b) Cooling on the TFIM with noise, where the noise takes the form of a local depolarizing channel with noise strength of $0.01$ on each qubit.}
\end{figure}

\subsection{Comparison}

An important result is that our numerical simulations confirm that the cooling protocol with annealing outperforms the cooling protocol without annealing on a test set. As shown in Figure 5, we select the one-dimensional transverse field Ising model as the cooling target, still using parameters $J_P=1, N_P=8$, with an initial state as a Neel state, and averaging over 30 samples. We randomly choose $g_P$, which varies from $1.2$ to $10$, resulting in systems with different energy level separations. Figure 5 (a) shows the differences among the three protocols. For smaller $J_{AP}^{max}$, due to too weak coupling strength, the cooling rate is very slow, and it fails to reach groud state in a short time. The influence of decay and other factors makes it impractical to set too small coupling strength in experiments. When $J_{AP}^{max}$ is larger, the initial cooling effects are similar between the annealed and non-annealed cooling protocols, but in the final converged state, the cooling protocol with annealing achieves much higher fidelity, while the other does not. If quantum simulation or quantum computing experiments require high precision in state preparation, opting for a non-annealed cooling protocol is clearly unreasonable, while the annealed cooling protocol can efficiently and accurately achieve cooling, making it a more robust option.

\begin{figure*}[htbp]
\centering
\includegraphics[width=\linewidth]{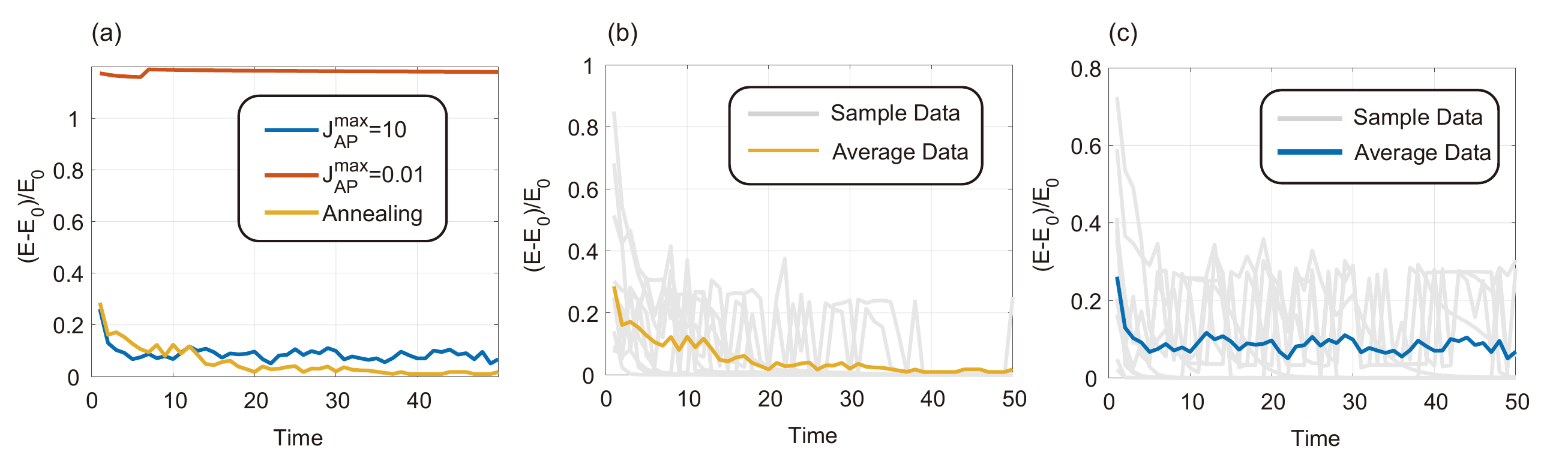}
\caption{(a) Testing effects of three different cooling protocols on a random test set. The blue curve represents the average cooling effect with a larger value of $J_{AP}^{max}$, the red curve corresponds to the average cooling effect with a smaller value, and the yellow curve represents the average cooling effect of the annealing protocol. (b) Average cooling curve of the annealing protocol along with several quantum trajectory samples from the cooling processes. (c) Average cooling curve of the cooling protocol with a larger value of $J_{AP}^{max}$ along with several quantum trajectory samples from the cooling processes.}
\end{figure*}

In Figures 5(b) and 5(c), we can see that the reverse transitions in the annealed cooling protocol gradually disappear, and the system evolves to converge to the ground state. However, the non-annealed protocol fails to achieve convergence to the ground state.

\subsection{Influence of types of noise}

At last, we demonstrate the influence on the cooling rate of types of noise, which includes $\sigma^x,\sigma^y,\sigma^z$ acting on qubits locally. As we know, if excitation caused by noise is corresponding to the interaction between the bath and the system, excitation can be removed efficiently. But other noise, especially topological excitation will be hard to be removed. As we demonstrate in figure 6, noise $\sigma^x$ has the least negative effects on cooling progress, as the other two types have lower fidelity than this one.

\begin{figure}[htbp]
\centering
\includegraphics[width=\linewidth]{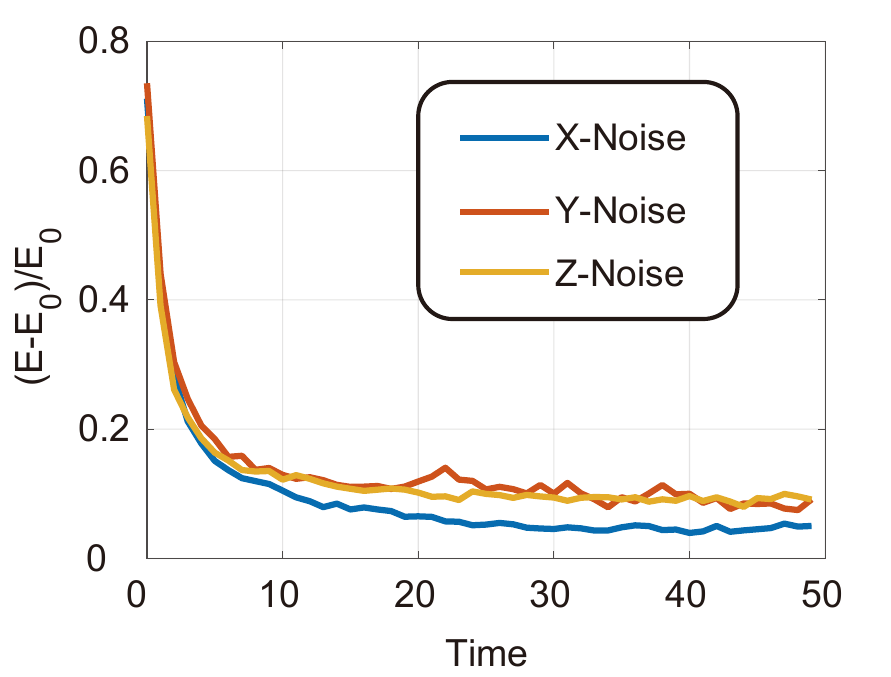}
\caption{The effect of different noise types on the cooling rate, only X-noise can be efficiently removed by the bath.}
\end{figure}


\section{Conclusion}

In this study, we proposed and analyzed a quantum cooling protocol based on time-modulated bath, which includes an annealing process, aiming at effectively preparing the ground state of quantum many-body systems

Firstly, by cooling the transverse field Ising model, we validated the effectiveness of the cooling protocol based on time-modulated bath. Numerical simulation results show that the energy of the system relative to the ground state is significantly reduced, demonstrating that the cooling protocol successfully drives the system close to the ground state. This result indicates that time-modulated Zeeman fields can effectively facilitate the evolution of the system, reducing the time required for cooling. Secondly, we compared the performance of annealed and non-annealed cooling strategies in the TFIM. When we use the annealed strategy, the cooling effect is significantly better than that of a fixed $J_{AP}^{max}$ strategy. This result suggests that gradually reducing the interaction strength between the bath and the system can effectively used in unknown or disorder system.

Considering that our proposed cooling protocol can still achieve good cooling results without knowledge of the Hamiltonian, this feature grants it broad applicability. Future studies may explore other types of quantum many-body systems and their applications in actual quantum computing platforms in complex environments. Moreover, this scheme suggests a possible way to cool a gapless system, which is difficult using adiabatic evolution\cite{polkovnikov2008breakdown}. However, quantum noise competing with cooling calls for interaction strength not too small, which is an obstacle when we cool a gapless system.

\appendix


\bibliography{apstemplate}

\end{document}